\documentclass[iicol,pdflatex,sn-mathphys-num]{sn-jnl}

\usepackage{graphicx}%
\usepackage{multirow}%
\usepackage{amsmath,amssymb,amsfonts}%
\usepackage{amsthm}%
\usepackage{mathrsfs}%
\usepackage[title]{appendix}%
\usepackage{xcolor}%
\usepackage{textcomp}%
\usepackage{manyfoot}%
\usepackage{booktabs}%
\usepackage{algorithm}%
\usepackage{algorithmicx}%
\usepackage{algpseudocode}%
\usepackage{listings}%
\usepackage{nicematrix}
\usepackage{rotating}
\usepackage{comment}
\usepackage{lineno}
\usepackage[capitalise,noabbrev]{cleveref}
\usepackage{siunitx}
\usepackage{hyperref}

\definecolor{orangeColour}{rgb}{1.0, 0.55, 0.0}

\definecolor{backcolour}{rgb}{0.8,0.8,0.8}

\lstdefinestyle{terminalCommands}{
    backgroundcolor=\color{backcolour},   
    basicstyle=\ttfamily\footnotesize\color{black},    breakatwhitespace=false,
    breaklines=true,                 
    captionpos=b,                    
    keepspaces=true,                                 
    showspaces=false,                
    showstringspaces=false,
    showtabs=false,                  
    tabsize=2
}

\begin{document}


\title[Sustainable Use of Computing Resources]{Enabling users to work sustainably on shared institute computing resources}


\author[]{\fnm{Niclas} \sur{Eich}}\email{niclas.eich@rwth-aachen.de}
\author[]{\fnm{Johannes} \sur{Erdmann}}\email{johannes.erdmann@physik.rwth-aachen.de}
\author*[]{\fnm{Martin} \sur{Erdmann}}\email{erdmann@physik.rwth-aachen.de}
\author[]{\fnm{Benjamin} \sur{Fischer}}\email{benjamin.fischer@rwth-aachen.de}
\author[]{\fnm{Paul} \sur{Gilles}}\email{paul.gilles@rwth-aachen.de}
\author[]{\fnm{Tim} \sur{Hauptreif}}\email{tim.hauptreif@rwth-aachen.de}
\author[]{\fnm{Jan} \sur{Kelleter}}\email{jan.kelleter@rwth-aachen.de}

\affil[]{\orgdiv{Physics Institute 3A}, \orgname{RWTH Aachen University},
\orgaddress{\street{Otto-Blumenthal-Stra{\ss}e}, \city{Aachen}, \postcode{52074}, 
\country{Germany}}}

\abstract{The VISPA project is a self-managed, mid-scale computing cluster that supports physics data analysis in research and teaching. Because the cluster is housed in a 1970s institute building with limited retrofit options, conventional efficiency upgrades would yield only minor energy savings. We therefore target sustainability primarily through user-centric measures. A monitoring system now records per-job energy consumption, while real-time data on the renewable share of the German power grid enable `green-window' scheduling. Users can query their individual energy consumption and carbon footprints, receive weekly reports, and tag jobs by project for aggregate accounting; memory records from previous runs help avoid oversubscription. All options are voluntary, fostering a cultural shift rather than imposing hard constraints. A simulation framework evaluates the potential impact of these measures. Together, the technological and behavioral interventions aim at medium- to long-term reductions in greenhouse-gas emissions by increasing resource awareness within the scientific community.}




\maketitle

\section{Introduction}
\label{sec:introduction}

The VISPA project is a prototype mid-sized institute computing cluster designed to meet local physics-data analysis needs in both research and teaching \cite{vispa}. Operated in a self-managed fashion, the VISPA cluster offers young researchers a broad playground for experimenting with new structures and ideas in shared computing. Numerous technological developments have been successfully implemented over the past 15 years and published. Three milestones attracted particular attention: first, a web-based user interface developed since 2010 that lets users obtain a desktop-like programming environment and computing resources with minimal effort \cite{Brodski:2011zz, Bretz:2012fu, Erdmann:2014vaa, vanAsseldonk:2015xng, Erdmann:2019aiy}, second the 2016 early deployment of $20$ GPUs dedicated to spread knowledge on training deep neural networks \cite{Erdmann:2018zsv}, and third fast plotting of terabytes of data using a caching system with SSD hard disks, which was developed in 2021 \cite{Eich:2022ctj}.

The VISPA project currently lists roughly 3,000 registered users, although only a small fraction makes continuous use of the cluster. The user base consists of roughly $50$ to $100$ students who, during the semester, tackle numerical homework aligned with their courses~--~ranging from short analyses of lab data to neural-network training~--~as well as about $15$ research-group members who access VISPA’s compute power directly for their scientific projects. Job orchestration is handled by the HTCondor batch system \cite{thain2005condor}.

Following a 2023 workshop on sustainability in the digital transformation of research on matter and the universe~\cite{Bruers:2023ftk}, the question of sustainability in VISPA’s operation received serious attention for the first time. Two main topics were discussed as the workshop: (1) technological advances for optimizing energy use, both in terms of compute performance and heat recovery, and (2) a cultural shift among scientists toward ensuring that the resources consumed in a physics data analysis remain in reasonable proportion to the expected scientific gain.

We would like to note here that high-performance data centers have been addressing similar issues for over 10 years and that there are numerous publications providing guidance on more sustainable computing. As an example from these publications, we mention the temporal coupling of compute jobs to the price of electricity \cite{Wiesner:2021shift, Haghshenas:2022shift}. As an example from physics, we note the effort of the \textit{Know your footprint} initiative, which enables an interactive estimation of greenhouse-gas emissions based on the activities of particle physicists \cite{Lang:2024wpo}.

Unlike modern high-performance data centers~--~many of which achieve outstanding PUE\footnote{PUE: Power Usage Effectiveness} ratings~--~the VISPA cluster resides in a building dating back to the 1970s, whose cooling system serves several other computing systems and for which the energy consumption could not be influenced or even determined. Such technological constraints leave very limited room for constructional efficiency upgrades, a situation typical for in-house departmental clusters. 
A major difference to a data center is that VISPA is not required to run at full load continuously. High demand occurs only intermittently, which provides flexibility in job scheduling.

Given these conditions, our primary opportunity for improving VISPA’s sustainability lies in fostering a cultural shift. We focus on ways to support the user community~--~via technical features~--~in increasing their awareness of resource usage and its balance against scientific benefit. Simultaneously, as cluster admins, we are committed to running the system on renewable energy wherever possible, thereby setting a positive example. Our overarching goal is a medium- to long-term reduction in greenhouse-gas emissions, achieved indirectly by raising awareness among our users.

This paper presents concrete improvement measures. First, we deployed an additional energy-monitoring setup. Power consumption reported by individual cluster components via on-board sensors is cross-checked by energy-metering power strips. We use these data to estimate the energy footprint of individual compute jobs. In addition, we query the Fraunhofer Institute for Solar Energy Systems ISE (Freiburg, Germany) for real-time and forecasted shares of renewables in the German power grid. The institute supplies a `traffic-light' indicator: green for a high share of renewable energy, yellow for a medium share, and red for predominantly fossil-based generation \cite{ISE_traffic_light}. We intend to provide equivalent local-grid information once it becomes available.

Several new features specifically target user awareness. All are opt-in; none are enforced. Users can:
\begin{itemize}
    \item Query their own energy consumption on the VISPA cluster for any time span;
    \item Choose sustainability options at job submission: run immediately, or defer execution until the share of renewables is high;
    \item Assign each job to a project category, allowing aggregate energy accounting at the project level;
    \item Display maximum memory usage in previously submitted jobs to support how much memory to request and hence ensure optimal sharing with other computing jobs.
\end{itemize}

Regarding the last point, users must specify an estimated memory requirement during job submission to enable efficient resource allocation, allowing multiple jobs to share a node and run in parallel when the combined memory usage fits within the node’s capacity. The request must exceed actual usage, yet should not reserve excessive capacity at the expense of other jobs. Since many submissions are repetitions with minor changes, VISPA displays  memory usage for similar past jobs, helping users pick a realistic value.

To assess the impact of all measures, we have built a flexible simulation framework. The following sections describe the VISPA compute cluster in detail and outline the technological developments that enable its more sustainable usage.

\section{VISPA computing cluster}
\label{sec:VISPA}

\begin{figure}[h]
    \centering
    \includegraphics[width=\linewidth]{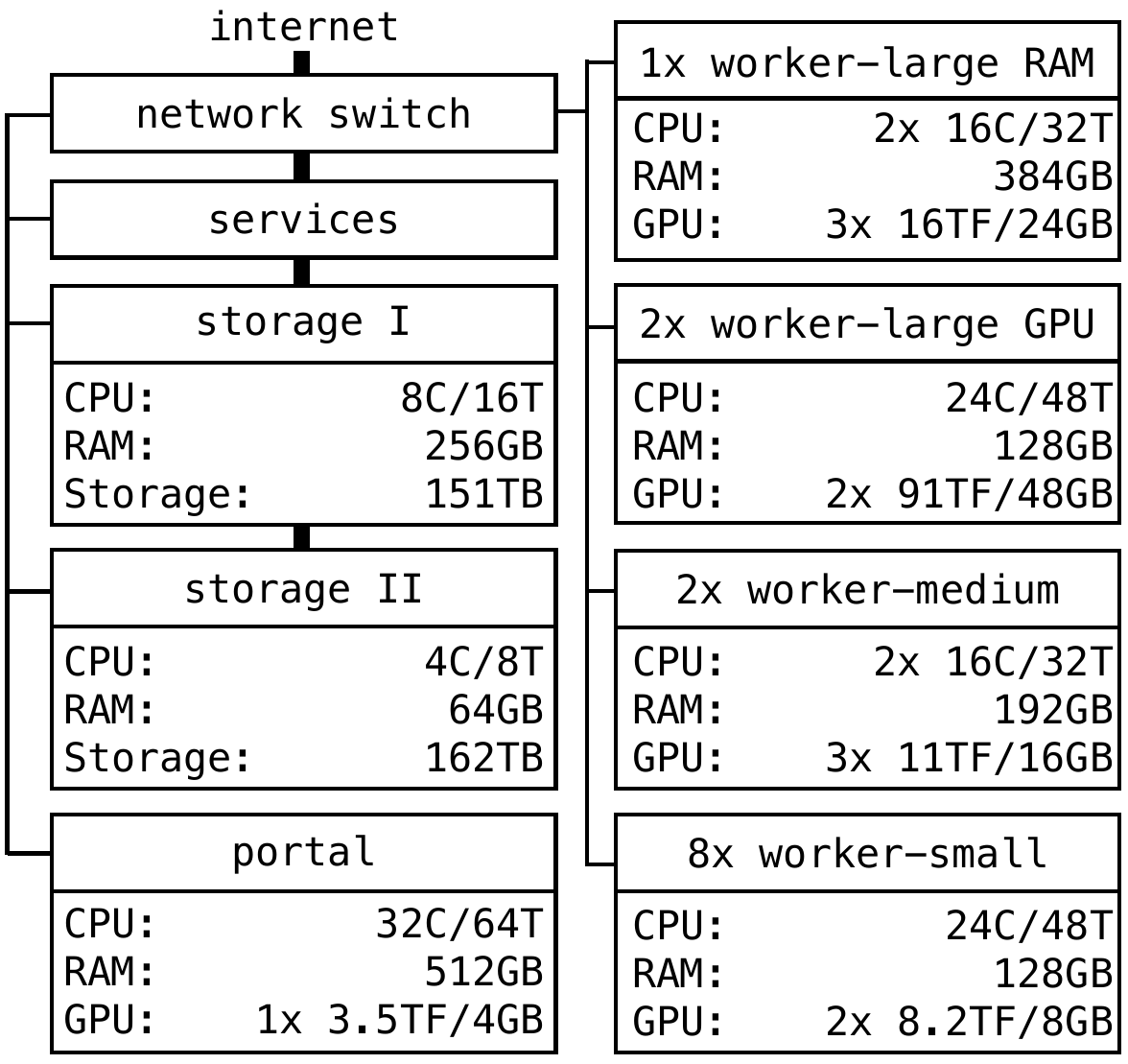}
    \caption{Computing resources of the VISPA cluster. VISPA has a shared storage system, 13 compute nodes and the service node \textit{portal} that supplies interactive log-in shells, a web-based desktop environment, and central batch-system control.}
    \label{fig:hardware_setup}
\end{figure}

The VISPA project's resources include a range of services, including a central access point (portal), 13 worker nodes with various computing power and memory capacity (CPUs and GPUs), storage management, and network functionalities (\cref{fig:hardware_setup}). A rather recent extensive description can be found in reference~\cite{Eich:2022ctj}.

\begin{figure}[h]
    \centering
    \includegraphics[width=\linewidth]{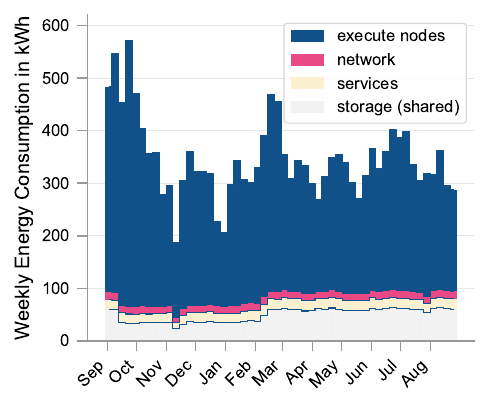}
    \caption{Total monitored energy consumption over one year (Sep24-Sep25) ($12$ of $13$ execution machines, referred to as worker in Figure~\ref{fig:hardware_setup}; one worker-medium was excluded due to PDU power limit).}
    \label{fig:total_energy_consumption}
\end{figure}

The first step toward a more sustainable operation of the VISPA cluster was the implementation of an energy consumption monitoring system. For this purpose, we use metered power distribution units\footnote{PDU: metered power distribution unit} (PDUs) that provide real-time measurements of power, voltage and current of nearly all components.
Due to the limited number of power ratings, monitoring was restricted to $12$ out of the $13$ worker nodes where one node classified as worker-medium was excluded (Figure~\ref{fig:hardware_setup}). An overview of the total power consumption over one year is shown in Figure~\ref{fig:total_energy_consumption}.

\section{Estimating energy per job}
\label{sec:job-energy}

\begin{figure}[h]
    \centering
    \includegraphics[width=\linewidth]{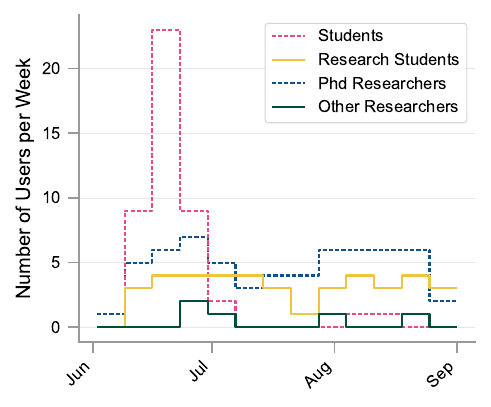}
    \caption{Weekly user counts are shown for students (working on exercises), research students (working on their thesis), doctoral researchers and other researchers (professors, postdoctoral researchers). Fluctuations occur primarily among students, driven mainly by the variations in their teaching schedules.}
    \label{fig:number_of_users}
\end{figure}

\begin{figure}[h]
    \centering
    \includegraphics[width=\linewidth]{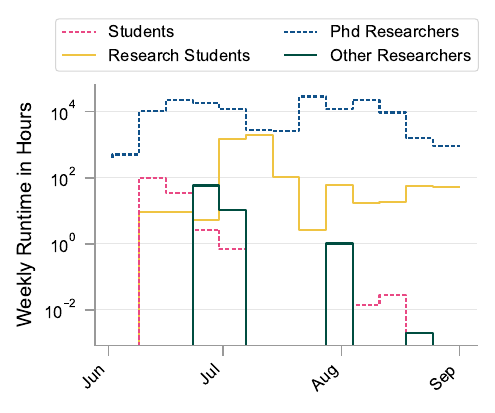}
    \caption{Total weekly runtime per user group. PhD students contribute the largest share of compute time despite not being the largest group; $100$ hours correspond to $\sim 4$ days and $1{,}000$ hours to $\sim 42$ days.}
    \label{fig:runtime_per_user_group}
\end{figure}

The users of the VISPA computing cluster comprise several groups: professors, postdoctoral researchers, doctoral researchers and students working on their thesis or on exercises related to lectures (Figure~\ref{fig:number_of_users}). 
Doctoral researchers are often regarded as the primary users, as they account for the majority of compute runtime (Figure~\ref{fig:runtime_per_user_group}). 

 Research at VISPA is primarily related to CMS analyses \cite{CMS:2008xjf} and the Pierre Auger Observatory \cite{PierreAuger:2015eyc}. Students engage with the cluster irregularly as part of their study programs, most often in introductory deep learning courses.

Compute time translates into energy use and environmental impact. To make this footprint visible and to encourage mindful use of shared resources, we estimate the energy consumption of each individual job and present it to the users. This per‑job, user‑facing feedback is intended to help balance scientific insight with responsible energy use. Users can access their data via command‑line tools, see Section~\ref{sec:PETRA}.

The system calculates and records per‑job energy estimates multiple times per day in a central database. In the following, the methodology used is described in detail.

We combine device‑level telemetry with metered PDUs to estimate per‑job energy. Specifically, we collect:
\begin{itemize}
  \item Total node energy \(E_{\text{machine}}\) measured by metered PDUs~\cite{gude_expert_power_control_8226, gude_expert_power_control_8035}.
  \item Total CPU energy \(E_{\text{cpu}}\), via RAPL\footnote{RAPL: Running Average Power Limit}.
  \item Job run interval $\Delta t_i$, defined as the time interval marking the start and end times of computing job $i$, also referred to as measurement interval.
  \item CPU usage per job $i$, \(R_{\text{cpu,i}}(\Delta t_j)\), as the average CPU usage percentage of the job’s cgroup\footnote{cgroup: Linux kernel feature} with respect to job $j$'s run interval.
  \item Total GPU energy \(E_{\text{gpu}}\) (via NVML\footnote{NVML: NVIDIA Management Library\label{footnote:NVML}}).
  \item Per‑process GPU compute utilization (SM\footnote{SM: Streaming Multiprocessor}) \(U_{\text{gpu}}\) (via NVML\textsuperscript{\ref{footnote:NVML}}).
  \item GPU energy per job \(i\), \(E_{\text{gpu},i}\), accumulated in increments by weighting total GPU energy \(E_{\text{gpu}}\) with per‑process utilization \(U_{\text{gpu}}\) (via NVML\textsuperscript{\ref{footnote:NVML}}) at maximum available resolution ($\SI{10}{Hz}$).
\end{itemize}

We break a job’s energy into GPU, CPU, and a residual \emph{overhead} component, which refers to non‑CPU/GPU hardware, such as main memory, fans, voltage regulators, and interconnects.

Starting with the CPU, CPU energy share of each job $i$ is determined proportionally based on its CPU utilization relative to all other jobs $j$ running concurrently on the same machine. Here, $R_{i}(\Delta t_i)$ denotes the average CPU utilization of job $i$ over the measurement interval $\Delta t_i$. The per-job CPU energy consumption is then given by
\begin{align}
    E_{\text{cpu},i} = 
    \frac{R_{i}(\Delta t_i)}{\sum_{j \in \text{machine}} R_{j}(\Delta t_i)} \, E_{\text{cpu}}.
\end{align}
The denominator sums over all jobs on the machine within the measurement time of job $i$. Jobs that start or end during the time window $\Delta t_i$ only contribute partially to the sum in the denominator. In the exclusive case (a single running job), that job receives the full CPU.

As overhead energy we refer to the machine‑level remainder:
\begin{align}
    E_{\text{overhead}} = E_{\text{machine}} - E_{\text{gpu}} - E_{\text{cpu}}.
\end{align}
The overhead energy is shared by the jobs running according to their runtime $t_j$ within the measurement interval $\Delta t_i$:
\begin{align}
    E_{\text{overhead},i} &= 
        \frac{t_{i}}
             {\sum_{j \in \text{machine}} t_{j}(\Delta t_i)}
        \, E_{\text{overhead}}.
\end{align}
The per‑job $i$ energy $E_i$ is then
\begin{align}
    E_{i} &= E_{\text{gpu},i} + E_{\text{cpu},i} + E_{\text{overhead},i}.
\end{align}

\begin{figure}[h]
    \centering
    \includegraphics[width=\linewidth]{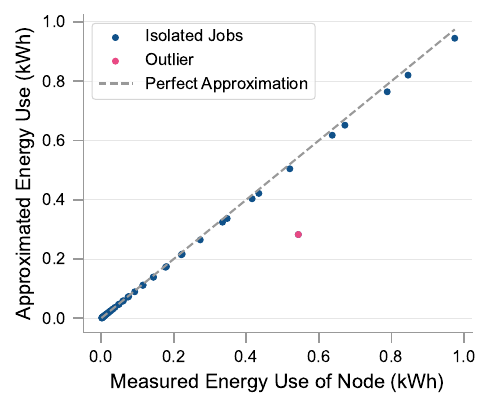}
    \caption{Isolated jobs: comparison of per‑job GPU energy (from per-process GPU compute utilization metric SM) with device‑reported total GPU energy. Deviations are small; transient underestimation can occur after idle periods.}
    \label{fig:isolated-jobs}
\end{figure}

As a cross-check, we focus on \emph{isolated jobs} that run exclusively on a single node, for which the job energy $E_{i}$ is expected to closely match the node’s total energy consumption $E_{\text{machine}}$. For \emph{isolated jobs} it is true that $E_{{\text{cpu},i}}= E_{\text{cpu}}$ and $E_{{\text{overhead},i}}= E_{\text{overhead}}$, while the GPU-related energy $E_{{\text{gpu},i}}$ only approximately equals $E_{\text{gpu}}$, since the GPU values obtained through the different approaches in the above-mentioned bullets. Although most measurements coincide well, we observe occasional outliers~--~one of which is highlighted in Figure~\ref{fig:isolated-jobs}~--~where the calculated energy significantly underestimates $E_{\text{gpu},i}$ and thus the total energy use. This underestimation occurs primarily for the first job after extended idle periods, due to limitations of the per-process GPU compute utilization metric (SM), which fails to fully capture short bursts of GPU activity following idleness. Importantly, deviations consistently correspond to underestimation rather than overestimation. We therefore inform users that the reported energy values should be interpreted as a lower bound on their actual consumption.

The overhead share represents baseline system consumption that cannot be uniquely tied to a single job. We report two views:
(1) distribute \(E_{\text{overhead}}\) proportionally to job runtimes (consistent with total machine energy), and
(2) assign the full baseline to each job (workstation‑style responsibility).
Each emphasizes a different fairness principle; our interface shows both as stacked components so that users can interpret the breakdown in context, see Figure~\ref{fig:cli_petra} in Section~\ref{sec:PETRA}.

For clarity, we calculate the user's energy consumption for execute machines exclusively (including, e.g., fans, memory) and do not include consumption outside the compute node (e.g., network infrastructure, shared file systems, or login nodes) in the per‑job accounting.

In sum, making energy visible at the job level is an actionable step toward a culture of mindful resource use, helping users consciously balance scientific progress with responsible energy consumption.

\section{Enabling sustainable resource usage}
\label{sec:sustainable}

In this section, we describe our access to information on renewable energy availability and the portfolio of tools provided to help users work more sustainably on the VISPA compute cluster.

\subsection{Renewable energy}
\label{sec:renew}

Time-shifting jobs to periods with higher renewable-energy availability is a valid strategy to reduce CO$_2$e emissions\footnote{Several greenhouse gases~--~including CO$_2$, CH$_4$, N$_2$O, HFCs, PFCs, SF$_6$, and NF$_3$~--~contribute to climate change. Their differing impacts are expressed relative to CO$_2$, and the combined effect is reported as CO$_2$-equivalent (CO$_2$e).}~\cite{GWD, Wiesner:2021shift} attributable to electricity use. The effectiveness of this strategy depends on the utilization level of the computing cluster; for instance, shifting jobs on fully utilized clusters offers limited potential for emission reduction. Given the varied and heterogeneous use of VISPA, this measure can be helpful.

\begin{figure}[h]
    \centering
    \includegraphics[width=1.0\linewidth]{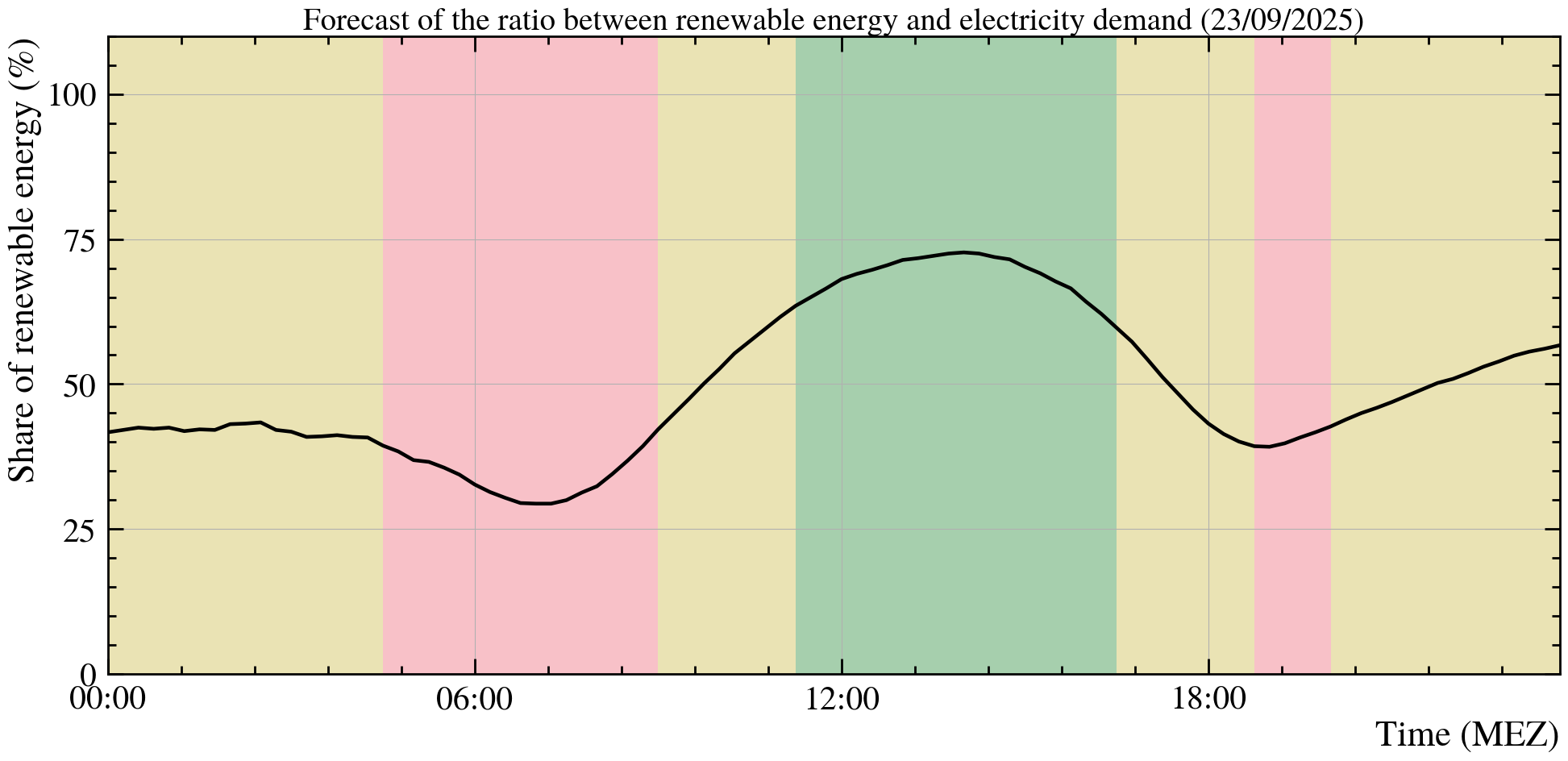}
    \caption{A visualisation of the traffic light metric provided by the Fraunhofer ISE \cite{ISE_traffic_light} as posted daily in internal communication channel.}
    \label{fig:trafficLight}
\end{figure}

To guide time-shifting decisions, we rely on data published by Fraunhofer ISE~\cite{ISE_traffic_light}. The dataset classifies the share of renewable energies in Germany’s electrical load using a traffic-light scheme (red, yellow, green) (Figure \ref{fig:trafficLight}). The signal is green if the renewable share either exceeds $95\%$ or is more than $10\%$ above the five-year monthly average. It is red if the share is more than $10\%$ below that five-year monthly average. All other values are categorized as yellow. Fraunhofer ISE publishes a next-day forecast each evening and updates it hourly thereafter.

Our HTCondor-based job management system queries the public Fraunhofer ISE API every two hours and stores the results in a JSON file. Based on this JSON, the current traffic-light status is propagated to the execute nodes by updating a ClassAd\footnote{ClassAd: data representation language used by HTCondor batch system} attribute every 15 minutes. The Class\-Ad mechanism in HTCondor is a flexible data representation language used to describe jobs, resources, and their requirements and properties within distributed computing environments~\cite{thain2005condor, raman1998matchmaking}. This external status informs scheduling decisions but only takes effect if the user opts in.

User control is provided through a sustainability flag. If the user does not specify this flag, the default value is red, ensuring that the user's job is not time-shifted. Users can set the flag either in their HTCondor submit description file or via a command-line argument of VISPA’s custom submission tool (see Listing~\ref{command:Sustainable}). The accepted values are red, yellow, or green (internally $0$, $1$, and $2$) and are stored as an attribute in the job’s ClassAd.

During matchmaking, HTCondor assigns a job to an execute node only if the node’s traffic-light value meets or exceeds the user’s requirement (i.e., job requirement $\leq$ machine status), in addition to satisfying the usual constraints on memory and other resources. Because renewable availability can be intermittent, stringent requirements may prolong queueing. To bound this delay, we introduce a maximum waiting time flag (second argument in Listing~\ref{command:Sustainable}), which allows users to specify a duration in hours for which the sustainability condition should be enforced. Once the specified time elapses, HTCondor disregards the sustainability constraint and matches the job as soon as all remaining conditions are met.
In practice this means that users can aim for a maximum degree of sustainability, while still knowing that their computations will be finished on their next working day or for their next presentation as long as enough computing resources are available and they respect their job's computation time.

\begin{lstlisting}[language=Bash, caption=Example terminal command to submit a job to the VISPA cluster using the sustainability and maxwait flags, label=command:Sustainable, captionpos=top]
submit --sustainable green 
       --maxwait 16 python job.py
\end{lstlisting}

\subsection{Energy per project}
\label{sec:energy_per_project}

To attribute compute resource usage to specific research activities, users can assign a project label to each submitted job. This label links jobs to their corresponding research projects, enabling project‑level attribution and analysis. The label is specified at submission time via the terminal; see Listing~\ref{command:Project}. Analogous to the time‑shifting preference, the project label is captured alongside the job metadata.

In conjunction with the per‑job energy estimator described in Section~\ref{sec:job-energy}, project labels enable aggregation of the total energy consumed by all jobs associated with a given project. These project‑level metrics can be reported in scientific publications and quantify a portion of the environmental impact attributed to the respective research effort.

\begin{lstlisting}[language=Bash, caption=Example terminal command to submit a job to the VISPA cluster using a project label, label=command:Project, captionpos=top]
submit --sustainable green 
       --project PaperXY python job.py
\end{lstlisting}

\subsection{User Communication via PETRA}
\label{sec:PETRA}

To enhance transparency and promote mindful resource use among VISPA cluster users, we developed PETRA (Programmed Energy Tracker for more Resource Awareness), a user-facing communication system provided both as an internal chat bot and as a terminal-based tool.

PETRA broadcasts daily forecasts of the expected share of renewable energy in Germany’s electrical load, using the Fraunhofer ISE traffic-light classification (red, yellow, green; see Figure~\ref{fig:trafficLight}). These signals enable power users to manually align their computing workloads with periods of higher renewable availability. While VISPA also supports automated scheduling to optimize renewable energy usage (Section~\ref{sec:renew}), PETRA’s lightweight, manual signaling mechanism offers a low-effort alternative that can be readily adopted on other computing infrastructures with minimal integration effort.

\begin{figure}[h]
    \centering
    \includegraphics[width=\linewidth]{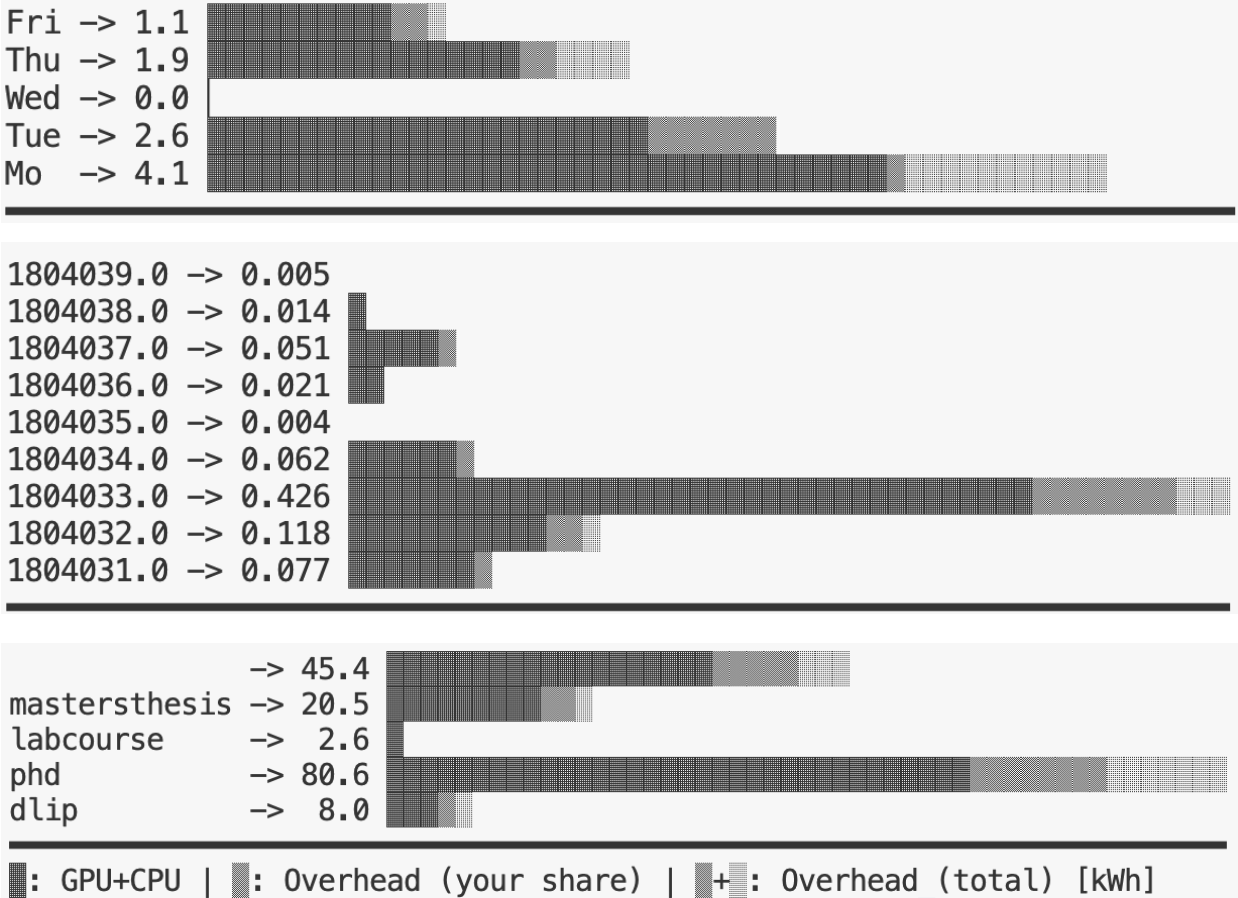}
    \caption{Screenshots of example outputs from the PETRA terminal interface. In addition to the total consumption figures, ASCII bar plots are used to visualize the energy consumption per weekday, per job, or per project (the latter requires the use of the project label introduced in Section~\ref{sec:energy_per_project}. Unlabeled jobs are categorized as one.)}
    \label{fig:cli_petra}
\end{figure}

Beyond grid forecasts, PETRA provides personalized estimates of individual energy consumption to increase awareness of users’ resource footprints (Section~\ref{sec:job-energy}). Access is via command-line tools that summarize energy usage over recent days, for recent jobs, and at the project level (Figure~\ref{fig:cli_petra}). The command-line interface is designed to be extensible and can accommodate additional statistical views in the future. Personalized consumption metrics were originally delivered exclusively via the internal chat bot, but migrating this functionality to the terminal has significantly improved accessibility, making PETRA available to all VISPA users independent of internal chat platform access.

By unifying these communication channels under the PETRA platform, we foster a culture of resource awareness and empower users to participate actively in sustainable computing practices.

\subsection{Memory requirement}
\label{sec:mem_req}

At VISPA, moderate cluster utilization often allows short wait times and enables shifting jobs and powering down idle machines as effective energy-aware strategies. To make these strategies broadly effective, we help users right‑size their resource requests~--~especially memory~--~so jobs fit more tightly and the scheduler can pack workloads more efficiently.

\begin{figure}[b]
    \centering
    \includegraphics[width=\linewidth]{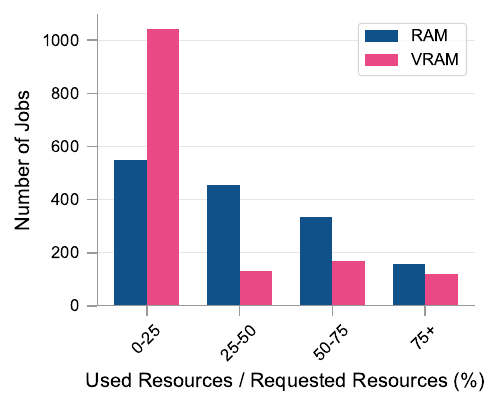}
    \caption{Comparison of requested versus actual used resources for RAM and VRAM, expressed as the ratio of used to requested values (\%). Less than half of the jobs utilize 50\% or more of their requested resources. The histogram shows jobs submitted on the VISPA cluster from June to the end of August 2025.}
    \label{fig:RequestedVSUsedResources}
\end{figure}

Typical schedulers (e.g., HTCondor) require users to declare memory needs. Because underestimation can cause job failures, users routinely add large safety margins, see Figure~\ref{fig:RequestedVSUsedResources}. 

\begin{figure*}[h]
    \centering
    \includegraphics[width=0.7\linewidth]{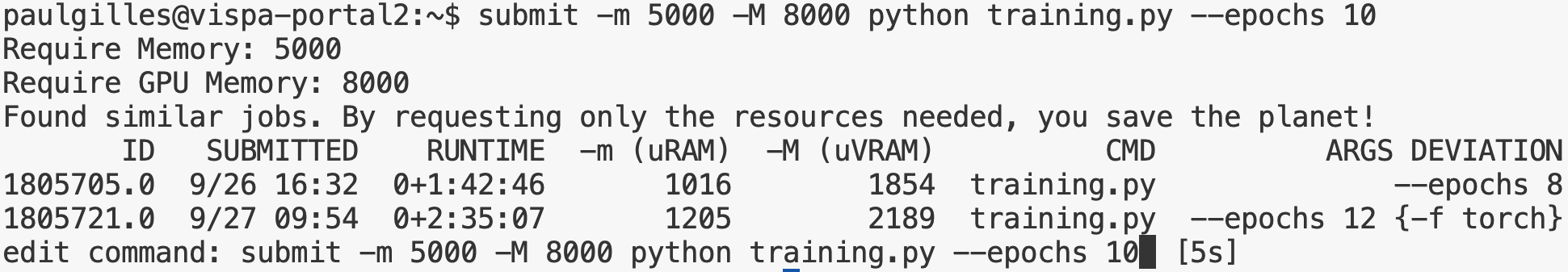}
    \caption{When submitting a job, the system searches the user's history for similar jobs. A job is similar if it is the same command and executable; arguments may vary.}
    \label{fig:memory_requirements_example}
\end{figure*}

\begin{figure}[h]
    \centering
    \includegraphics[width=1.0\linewidth]{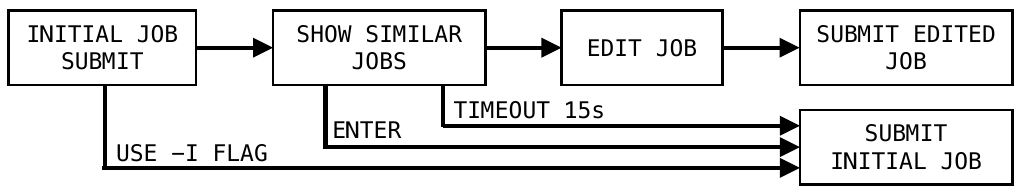}
    \caption{Flow diagram of the terminal tool illustrating its unobtrusive design.}
    \label{fig:memory_requirements_schematic}
\end{figure}

The result is over‑provisioned jobs that block resources they do not actually use, leading to longer queues and lower overall efficiency. While predicting requirements in advance is hard, many users submit the same or similar jobs repeatedly, and past runs provide valuable insights.

We therefore introduce a lightweight command‑line prompt that, just before submission, shows statistics from previous similar jobs, including their actual resource usage (Figure~\ref{fig:memory_requirements_example}). 

Users can adjust their requests based on this evidence. The workflow remains unchanged: jobs auto‑submit after a short countdown, pressing Enter confirms immediately, and the feature can be fully disabled for automated pipelines (e.g., Dask). A complete schematic is shown in Figure~\ref{fig:memory_requirements_schematic}.

In short, data‑informed, right‑sized requests yield denser scheduling, shorter queues, and less energy waste~--~advancing the broader goal of a culture of mindful, responsible use of shared computing resources.

\section{Simulation method of resource usage}
\label{sec:simulation}

We developed a digital twin of the VISPA computing cluster to simulate and compare sustainability interventions, enabling quantitative analysis of trade-offs between environmental impact and user-perceived quality of service. The twin can replay historical VISPA workloads or execute synthetic job mixes at an accelerated time scale, i.e., faster than real time. From the users’ perspective, perceived quality of service is dominated by queue waiting times and job runtimes, whereas the sustainability of the cluster is characterized by its total electricity consumption and the associated CO$_2$e emissions.

\subsection{Simulation Setup}

To mirror the production environment, the digital twin is built from Docker containers running unmodified HTCondor instances. Each container emulates a single VISPA execute node and uses the original configuration files from the production cluster. Scheduling decisions, matchmaking, and policy enforcement are thus governed by the same software stack as in production, and modifications validated in the twin can be ported back to the real cluster with minimal effort.

The twin itself runs on a single desktop workstation, but each HTCondor instance is initialized with artificial resource capacities to emulate larger cluster scales. Synthetic numbers of CPU cores, system memory, and GPUs per node allow us to evaluate scheduling and sustainability policies under realistic high-load conditions without a full-size physical cluster. To accelerate experimental cycles, the simulator advances its internal time by a factor of ten relative to wall-clock time: job runtimes and queueing delays are compressed accordingly, while the underlying scheduling logic remains unchanged. All reported energy metrics are rescaled to correspond to real-time operation.

Because the containers run plain HTCondor, simulated jobs can express the same constraints and ClassAd attributes as on the real cluster. For experimental clarity, we restrict job requirements to a core set of attributes: the number of CPU cores, system memory (RAM), the number of GPUs, GPU memory (VRAM), and a sustainability preference flag used by the scheduler to prioritize energy-aware placements as described in Section~\ref{sec:renew}. During execution, HTCondor reserves the requested resources for the job’s runtime and blocks other jobs from using them, reproducing the allocation behavior of the production system. Since requested resources often exceed actual resource usage, we additionally inject the measured real resource consumption into the simulation instead of relying on the requested allocations. 

Obtaining fully time-resolved per-job metrics in production would require continuous live tracking with nontrivial overhead, so we rely on constant-over-time summaries. For every job, HTCondor provides peak system memory and peak GPU memory, as well as average CPU-core and GPU utilization over the job’s lifetime. These quantities are tracked during execution and written to the job history upon completion. We ingest these values as if they were constant over the job duration and superimpose them across all concurrently running jobs on a given machine to reconstruct per-machine resource occupancy. A scraper runs periodically to gather the per-machine usage. While this approximation discards fine-grained temporal structure, we expect the resulting inaccuracies to average out across the large number of jobs observed.

In the final stage of the pipeline, the simulated per-machine resource traces are passed to a deep learning model. At each simulation time step, the model estimates the instantaneous power draw from the corresponding resource occupancy. Integrating the predicted power over simulated time yields estimates of the total energy consumption of each workload and of the cluster as a whole. In parallel, we log queueing times for all jobs in the twin to quantify the impact of sustainability-oriented scheduling policies on user-perceived latency and throughput.

\subsection{Calculation of the Energy Usage}

\begin{figure}[h]
    \centering
    \includegraphics[width=0.5\linewidth]{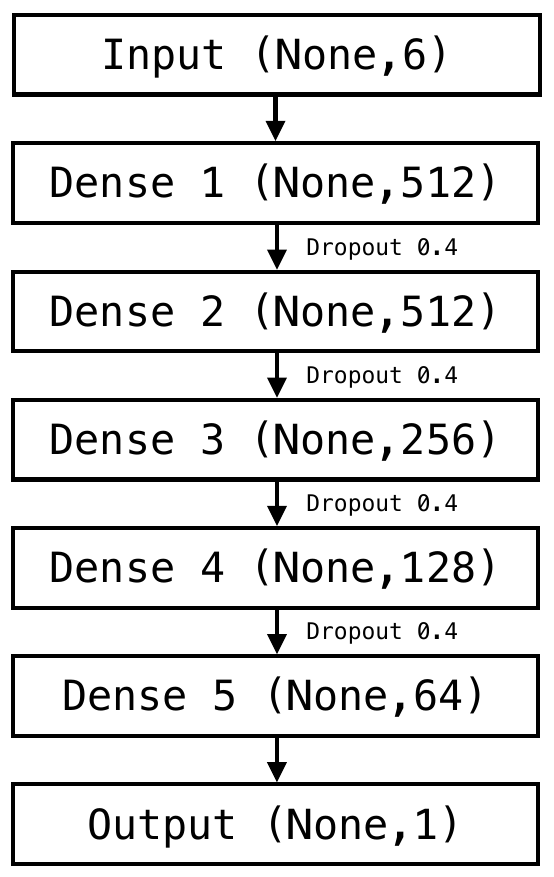}
    \caption{Architecture of the deep neural network used to convert simulated resource usage into estimated power consumption. The model consists of five fully connected hidden layers with nonlinear activations and dropout regularization; layer annotations indicate the dimensionality of the outputs.}
    \label{fig:sust_dnn}
\end{figure}

To estimate instantaneous power draw from resource usage, we employ a deep neural network. The architecture is shown in Figure~\ref{fig:sust_dnn}. The network consists of five fully connected hidden layers with nonlinear activations and dropout regularization to mitigate overfitting and improve generalization to unseen workloads. Categorical attributes such as the machine identifier are encoded appropriately.

The training corpus comprises seven weeks of production cluster history. Resource usage was logged on each machine at 1-minute intervals and paired with the corresponding power measurements from the rack-level power distribution unit (PDU), as described in Section~\ref{sec:VISPA}. At each timestamp, the feature vector contains the time-averaged CPU and GPU utilization across all jobs on the machine, the machine identifier, the total number of active jobs, the number of jobs using at least one GPU, and the accumulated runtime of the GPU-using jobs. The latter feature captures the common pattern that GPU workloads front-load CPU activity (e.g., for data loading or model initialization), which is not reflected in time-averaged utilizations alone.

Most machine-timestamp combinations in the raw log correspond to idle periods with no CPU and no GPU usage. To reduce the resulting class imbalance and focus training on informative examples, we randomly discard 90\% of entries with zero CPU and zero GPU utilization, leaving $244,025$ samples. One machine lacked PDU monitoring during the logging period and is therefore excluded from training; because it is hardware-identical to a monitored node, it is mapped to the corresponding model during evaluation.

\begin{figure}[h]
    \centering
    \includegraphics[width=\linewidth]{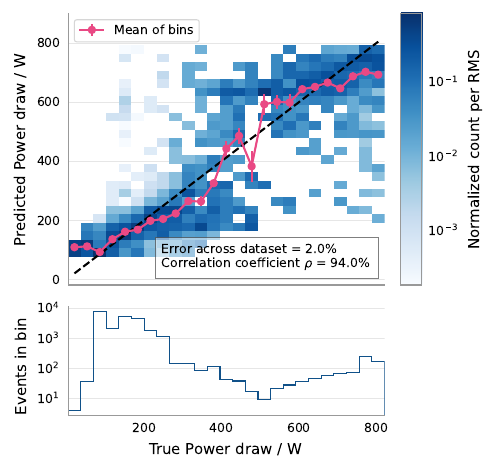}
    \caption{Top: predicted versus true power usage for the test set, as measured at the PDU. The two-dimensional histogram is normalized along the axis of the true power. Bottom: distribution of true power values in the test data. The model achieves 94\% correlation and 2\% total uncertainty.}
    \label{fig:dnn_perf}
\end{figure}

We split the dataset into training, validation, and test subsets using an 80/10/10 partition. Model quality is summarized in Figure~\ref{fig:dnn_perf}, which shows the predicted power versus the true PDU measurements on the test set. The two-dimensional histogram is normalized by the number of events in each bin along the truth axis. The network achieves a Pearson correlation coefficient of 94\% and an aggregate error of 2\% over the test subset. The latter is computed by summing predictions and ground-truth power across all test samples and comparing the totals; this aggregate reporting format is used throughout the following chapter to match the energy-centric perspective of our study.

\section{Investigations on effects of sustainability measures}
\label{sec:fossil}

We use the simulation framework described in Section~\ref{sec:simulation} to evaluate sustainability measures on the VISPA cluster and to quantify their effects on environmental impact and user experience. Some measures can be implemented by administrators, while others require explicit user action.

For each scenario, we record job queueing times and power usage as described in Section~\ref{sec:simulation}. CO$_2$e emissions are computed by weighting the instantaneous power draw with historical data on the composition of the German electricity mix \cite{ISE_historic_data} and applying generation-method-specific average CO$_2$e emissions per kWh \cite{IPCCReport}. The results should be interpreted as trends rather than universal numbers, as absolute gains depend on the local energy mix, workload characteristics, and cluster hardware.

\subsection{Description of sustainability measures}

We investigate three distinct, combinable sustainability measures, each implemented in an aggressive, idealized variant to expose its maximum potential impact.

\paragraph{A. Better resource requests}

As discussed in Section~\ref{sec:mem_req}, most users substantially overestimate the resource needs of their jobs, particularly for memory. VISPA offers tooling to provide feedback on actual usage, but perfect alignment is rare. In the simulation, we nonetheless assume that users predict their jobs’ resource usage exactly, providing an upper bound on the benefits attainable through improved guidance.

\paragraph{B. Dynamic machine shutdown}

Because large portions of the VISPA cluster are idle for much of the day, suspending or powering down machines during idle periods can save energy. We employ a simple policy: a machine is suspended if it has not been assigned a job for at least five minutes and is activated when a new job is assigned while it is suspended. The shutdown condition is evaluated at five-minute intervals to emulate aggressive capacity adaptation.

\paragraph{C. Sustainability-aware scheduling}

To exploit temporal variations in the carbon intensity of the electricity mix, we use the sustainability flag introduced in Section~\ref{sec:renew}, driven by an ``energy traffic light'' indicating green, yellow, or red periods. For the simulations here, jobs are allowed to start only during green periods, and users are assumed to tolerate unlimited waiting. This illustrates what could be achieved if sustainability were the dominant optimization criterion.

\subsection{Evaluation of an academic workload scenario}
\label{sec:hists}

To validate the implementation of the measures and compare against intuitive expectations, we construct a tightly controlled artificial scenario based on a real VISPA job. The job parameters are summarized in Table~\ref{tab:simple_job}. Runtime cannot be estimated reliably at submission, and average GPU utilization is not requested explicitly; instead, the VRAM request determines GPU allocation, as is common among VISPA users.

In this scenario, the example job is submitted $2,000$ times at $t = 0$ and another $2,000$ times at $t = \SI{12}{h}$. For configurations involving the energy traffic light, the signal is red for the first three hours, blocking all job execution, and green for the remainder of the day. For the green/red energy composition and CO$_2$e per kWh, we use the respective 2024\footnote{Data for 2025 are incomplete due to outages in the traffic-light monitoring.} annual averages of the German electricity mix. All simulations run for one day, with parameters chosen such that all jobs finish within \SI{24}{h}.

\begin{table}[h]
\centering
\begin{tabular}{|l|c|c|}
    \hline
     & Requested & Used \\ \hline
    Runtime (s) & - & $534$ \\ \hline
    Average CPU (cores) & $16$ & $6$ \\ \hline
    Memory (MB) & $8,000$ & $1,023$ \\ \hline
    Average GPUs & - & $0.2$ \\ \hline
    VRAM (MB) & $8,000$ & $708$ \\ \hline
\end{tabular}
\caption{Parameters of the job used in the artificial scenario, motivated by a realistic job submission. Where possible, a distinction is made between requested and used resources. Runtime cannot be estimated at submission, and average GPU utilization is not requested explicitly; instead, the VRAM request governs GPU allocation.}
    \label{tab:simple_job}
\end{table}

\begin{figure*}[h]
    \centering
    \includegraphics[width=0.9\linewidth]{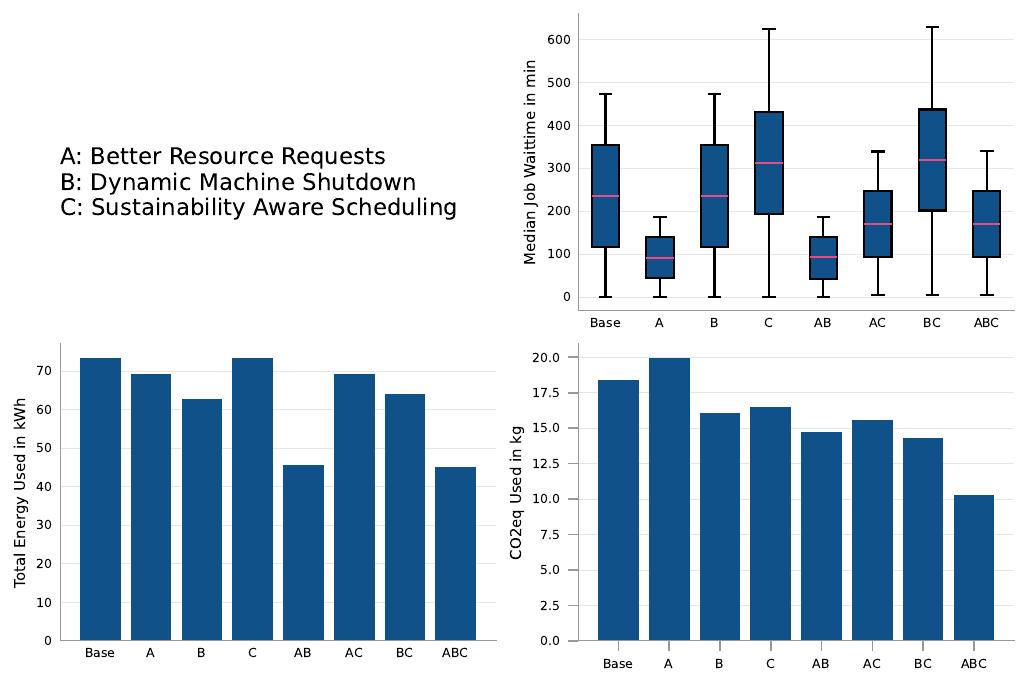}
    \caption{Simulation results for the academic workload scenario. Evaluation metrics are median waiting time, total energy usage, and total CO$_2$e emissions. More accurate resource requests substantially decrease waiting times while slightly reducing energy usage. Dynamic machine shutdown reduces power usage with negligible impact on waiting times. Sustainability-aware scheduling shifts execution into greener periods, significantly reducing CO$_2$e emissions at the cost of increased waiting times.}
    \label{fig:plot_ac}
\end{figure*}

The artificial case serves to validate the simulations and checks whether the observed effects match naive expectations. The results are summarized in Figure~\ref{fig:plot_ac}. We focus on three metrics: median job waiting time, total energy usage, and total CO$_2$e emissions.

Applying the measures individually yields outcomes largely consistent with intuition. More accurate resource requests (measure A) reduce waiting times substantially because jobs can be packed more tightly onto the available hardware. Comparing the requests in Table~\ref{tab:simple_job} with the VISPA configuration in Figure~\ref{fig:hardware_setup} suggests a packing factor between $2.2$ and $4$, depending on node configuration, consistent with waiting time reduction. Power usage decreases slightly relative to the baseline, but CO$_2$e emissions increase, as more jobs are executed during the initial three-hour red period when the share of fossil fuels in the grid is higher.

Dynamic machine shutdown (measure B) leaves the median waiting time essentially unchanged in this idealized setting. Wake-up checks are aligned with job distribution, and power-on is modeled as instantaneous, so additional latency is negligible. Power usage and CO$_2$e emissions are reduced because idle runtime is minimized by suspending unused machines.

Sustainability-aware scheduling (measure C), which restricts job start times to green periods, increases the median waiting time by roughly \SI{90}{min}. This is compatible with the three-hour initial red phase, since approximately half of all jobs must wait until $t = \SI{3}{h}$ before any are allowed to run. Total energy usage changes little relative to the baseline, but CO$_2$e emissions decrease because execution is shifted to greener periods.

Combining the measures amplifies many of their benefits. The combination of better resource requests with dynamic shutdown (A+B) yields strong reductions in power usage at negligible additional waiting time: tighter packing extends idle periods, which in turn are exploited more effectively for shutdowns. Combining better requests with sustainability-aware scheduling (A+C) reduces CO$_2$e emissions at nearly unchanged total energy consumption by shifting computation to greener periods while keeping waiting times below the baseline. Adding sustainability-aware scheduling to dynamic shutdowns (B+C) primarily lowers CO$_2$e by moving shutdown phases into intervals with high fossil fuel utilization. 

When all three measures are applied together (A+B+C), the system reaches the lowest total energy consumption and CO$_2$e emissions across all configurations, while preserving median waiting times below those of the baseline scenario.

\subsection{Evaluation using a realistic workload scenario}

We also evaluate a realistic scenario consisting of $11$ consecutive days of VISPA job history, from late July to early August 2025. During this period, a total of $9{,}768$ jobs were submitted, and all resource usage and energy composition data for the simulation are taken from the same time span.

This slice of history is characterized by very low GPU utilization: only about 4\% of jobs request any GPU resources, and only 0.4\% use more than 10\% GPU on average, so the cluster is mostly CPU-bound. The energy traffic light is based on a rolling monthly average of the previous five years; in 2025 this reference still includes the full COVID-19 period with reduced electricity demand, leading to more frequent and longer red phases. During this period, machines are idle about 90\% of the time on average, i.e., the cluster is strongly underutilized.

\begin{figure*}[ht]
    \centering
    \includegraphics[width=0.9\linewidth]{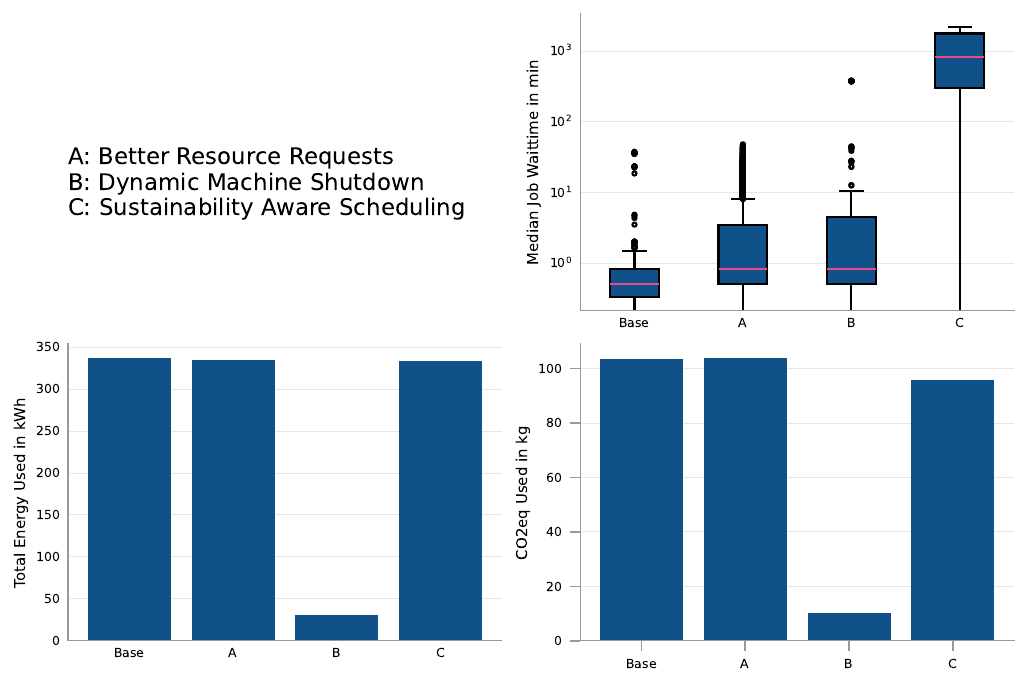}
    \caption{Simulation results for the realistic case. In contrast to naive expectations, better resource requests increase median waiting times, which we attribute to artifacts of slot partitioning and missing job-dependency information in the historical log. Dynamic machine shutdown delivers large reductions in power usage and CO$_2$e emissions on this underutilized cluster. A naive implementation of sustainability-aware scheduling, in which jobs wait indefinitely for green periods, further reduces emissions but leads to unacceptably long waiting times; a capped waiting period is therefore recommended in practice.}
    \label{fig:plot_real}
\end{figure*}

We apply the three measures to this realistic workload, with results shown in Figure~\ref{fig:plot_real}. 
In this setting, better resource requests (measure A) behave counter to naive expectations: power and CO$_2$e usage remain essentially unchanged, while median waiting times increase. The main reason is a simulation artifact from slot partitioning and job grouping: in production, many submissions are clustered via tools such as \emph{coffea}~\cite{CMS:2020kpn,Gray:2023ajb}, which assign identical, averaged resource requests and hide inter-job dependencies from the HTCondor history. Retrofitting per-job “perfect” requests makes the workload more heterogeneous; combined with fixed slot sizes and missing dependency information, this causes fragmentation, where small jobs fill gaps and larger jobs wait longer. This is a limitation of our simulation and we do not expect similar behaviour on VISPA. In this low-utilization regime, idle power dominates and tighter packing does not reduce the number of active machines, so measure A yields no net sustainability benefit unless coupled to capacity-management mechanisms (e.g., dynamic shutdown).

Dynamic machine shutdown (measure B) increases the median waiting time to just under one minute, consistent with the waiting period between job matching cycles\footnote{HTCondor negotiating cycles}: with an average utilization of only 10\%, many jobs trigger a machine wake-up before starting, forcing them to wait an additional cycle. Power usage and CO$_2$e emissions are reduced by roughly 90\%, matching the average downtime fraction of the machines. This reduction appears large, as only idle power should be saved. However, due to low GPU utilization, most jobs operate only slightly above idle power and are largely CPU-bound. Moreover, the mean curve in Figure~\ref{fig:dnn_perf} shows that the power model overestimates the lowest power draws, biasing idle predictions toward pure CPU-load levels. Even accounting for this bias, dynamic shutdown remains the most effective lever for reducing unnecessary energy consumption and greenhouse gases under typical VISPA loads.

Sustainability-aware scheduling (measure C) only yields modest reductions in CO$_2$e emissions. However, it is important to note that the sustainability-aware scheduling exclusively affects the CO$_2$e emissions resulting from computational workload, as the power consumption of idling machines can not be time-shifted. Thus, the maximal reduction in CO$_2$e emissions that can be achieved with sustainability-aware scheduling is given by the CO$_2$e emissions remaining after measure B is employed. Comparing measure B with the difference between the baseline and measure C in Figure~\ref{fig:plot_real} indicates that the sustainability-aware scheduling has a substantial impact given the small potential for reductions of the CO$_2$e emissions. In a scenario with higher cluster utilization, in which the energy consumption is not dominated by idling machines, sustainability-aware scheduling can be expected to be more effective.  

The modest reductions in CO$_2$e emissions come at the cost of very high waiting times, sometimes exceeding a full day. This shows that a naive ``wait until green'' policy is unsuitable for production use. We instead recommend imposing a maximum waiting period, as in Section~\ref{sec:renew}, and combining sustainability-aware scheduling with other measures such as better resource requests and machine shutdown to keep delays acceptable.

Because the 2025 traffic light compares current demand against a five-year reference that includes the COVID-19 period with atypically low consumption, red and yellow phases are more frequent and prolonged than in a typical year. The waiting times observed here should therefore be interpreted as a worst-case illustration rather than a typical outcome.

\section{Accomplished initiatives and insights gained}
\label{sec:Conclusion}

Our work was motivated by the white paper on sustainability in the digital transformation of research on matter and the universe~\cite{Bruers:2023ftk}. We operate an institute-level compute cluster that provides substantial resources for research and education. Here, we examine how sustainability measures can be implemented in practice on an existing cluster located in a roughly 50-year-old building with a pre-existing cooling system serving several other computing systems and for which the energy consumption could not be determined. The embodied CO$_2$e associated with the construction of the building and the already installed hardware is effectively fixed and, like the cooling system, was not considered further in this study.

Within this scope, we focused on several measures designed to enable users to utilize the cluster more sustainably. In addition, we investigated options for partially powering down the cluster.

The first set of measures concerns the services we provide to users: detailed reporting of energy consumption per user, per job, and per project. To this end, we compared on-device energy readings with externally measured power consumption at the device level and calibrated the on-device values accordingly. The resulting calibrated consumption values reach an accuracy at the 10\% level or better. Project-level reporting requires users to provide a project identifier when submitting jobs.

Users are also given the option to schedule job start times for periods that are more favorable from a sustainability perspective, as indicated by an electricity “traffic light” signal~\cite{ISE_traffic_light}. A point to note here is that the traffic-light colors are set dynamically, relative to a monthly average computed over multiple years, which still includes pandemic periods and which is associated with a different energy mix due to the lower energy consumption. Users can specify a maximum waiting time so that results are available on time, regardless of the traffic-light status.

Finally, we take job memory requirements into account in order to maximize the number of jobs that can be executed concurrently on a single machine. Users’ memory requests are often heavily overestimated to ensure that jobs do not run out of memory and finish successfully. In our approach, users receive feedback before submission about the actual memory usage of their previous jobs and can therefore provide an optimized memory request. This is particularly helpful during the development of data analyses, where repeated bug fixes often require rerunning identical jobs.

To evaluate the different measures, we developed a digital twin of the compute cluster that reproduces its power consumption and the associated CO$_2$e emissions. This framework also allowed us to study scenarios in which the cluster is only partially active, with all idle workers powered down.

Using this simulation, we tested two scenarios. The first was an artificial, research-intensive workload consisting of several thousand similar jobs, submitted in two batches on a single day, assuming a typical sequence of the electricity traffic-light signal. The second scenario was derived from a eleven-day period of actual cluster usage.

In the research-intensive scenario, the simulation showed that the combination of all three measures~--~A) realistic memory requests, B) dynamic shutdown of idle workers, and C) aligning job start times with the green phases of the electricity traffic light~--~yielded the largest reduction in CO$_2$e. In contrast, when replaying a real usage period with relatively low overall cluster utilization, powering down idle workers offered by far the greatest potential for reducing CO$_2$e emissions.

Overall, our measures enable users to operate the compute cluster in a more sustainable manner. Along the way, we learned that the challenges lie in quite some details. In particular, the heterogeneity of usage times and workloads~--~from simple Python scripts or trainings of small neural networks by students, to week-long trainings of deep neural networks or the simultaneous submissions of $\mathcal{O}(\num{10000})$ jobs by researchers~--~leads to very diverse requirements that, in turn, strongly influence the usefulness and impact of each individual sustainability measure. By systematically exposing resource and energy footprints at the level of users, their jobs, and projects, our approach equips users with means to identify which measures are most effective for their specific workflows and to integrate sustainability considerations into their everyday computing practice. We hope that this combination of infrastructure mechanisms and transparency towards users can serve as a template for other academic computing environments seeking to reduce their climate impact.
\backmatter

\bmhead{Supplementary information}
More information and examples regarding the implementation of our measures can be found in a set of public GitLab repositories: \\
Sustainability-aware scheduling in HTCondor:\\\url{https://git.rwth-aachen.de/3pia/vispa/sustainability/sustainability-flag-publication}\\
Per-Job Energy Approximation at VISPA: \\\url{https://git.rwth-aachen.de/3pia/vispa/sustainability/energy-per-job-publication} \\
Terminal interface for more accurate resource requests (Better Resources): \\\url{https://git.rwth-aachen.de/3pia/vispa/sustainability/better-resources-publication}\\
Terminal interface showing energy consumption as stacked bar plots (PETRA in terminal): \\\url{https://git.rwth-aachen.de/3pia/vispa/sustainability/petra-in-terminal-publication}\\
VISPA simulations (Mini-Vispa): \\\url{https://git.rwth-aachen.de/3pia/vispa/sustainability/mini-vispa-publication}

\bmhead{Acknowledgements}

We wish to thank P.~Wissmann and F.~Zinn for numerous helpful discussions.
This work is supported by the Ministry of Innovation, Science, and Research of the State of North Rhine-Westphalia, as well as by the Federal Ministry of Research, Technology, and Space (BMFTR) in Germany. Language translation support was received by DeepL and Open AI GPT5.

\section*{Declarations}

The authors have no competing interests to declare that are relevant to the content of this article.

\bibliography{sustainability.bib}

\end{document}